\documentclass[runningheads]{llncs}
\usepackage[T1]{fontenc}
\usepackage{graphicx}

\usepackage[hidelinks]{hyperref}
\usepackage{amssymb} 
\usepackage{xspace}
\usepackage{tikz}
\usepackage{amsmath}
\usepackage{listings}
\usepackage[disable]{todonotes}
\usepackage[inline]{enumitem}
\usepackage{subcaption}
\usepackage{adjustbox}
\usepackage{booktabs}
\usepackage{tabularx}
\usepackage{array}
\usepackage{makecell}
\usepackage{wrapfig}
\usepackage{multirow}

\usetikzlibrary{arrows.meta, positioning, fit, calc, shapes.geometric, backgrounds, matrix}

\usepackage{color}
\usepackage{xcolor}

\newcommand{\cfile}{\texttt{.c} file\xspace}
\newcommand{\hfile}{\texttt{.h} file\xspace}
\newcommand{\isfile}{\texttt{.is} file\xspace}
\newcommand{\cfiles}{\texttt{.c} files\xspace}
\newcommand{\hfiles}{\texttt{.h} files\xspace}

\newcommand{\plugin}{VerNFR\xspace}
\newcommand{\tool}[1]{\textsc{#1}\xspace}
\newcommand{\cppchecker}{\texttt{Cppcheck}\xspace}
\newcommand{\agent}{agent\xspace}
\newcommand{\agents}{agents\xspace}
\newcommand{\module}{module\xspace}
\newcommand{\Module}{Module\xspace}

\newcommand{\framac}{\tool{Frama-C}}
\newcommand{\wpplugin}{\tool{Wp}}
\newcommand{\wprteplugin}{\tool{Wp-rte}}
\newcommand{\eva}{\tool{EVA}}
\newcommand{\aorai}{\tool{aoraï}}
\newcommand{\metacsl}{\tool{Met\-ACSL}}
\newcommand{\gcc}{\tool{GCC}}
\newcommand{\acsl}{ACSL\xspace}
\newcommand{\toolchain}{toolchain\xspace}

\newcommand{\Toolchain}{Toolchain\xspace}
\newcommand{\interfacespec}{interface specification\xspace}
\newcommand{\interfacespecs}{interface specifications\xspace}
\newcommand{\iscontract}{IS contract\xspace}
\newcommand{\iscontracts}{IS contracts\xspace}
\newcommand{\autodeduct}{\tool{AutoDeduct}}

\newcommand{\nonterminal}[1]{\mbox{$\langle #1 \rangle$}}
\newcommand{\entrypoints}{\mbox{$\mathit{EntryPoints}$}}
\newcommand{\entryorder}{\mbox{$\mathit{EntryOrder}$}}
\newcommand{\extmodule}{\mbox{$\mathit{ExtModule}$}}
\newcommand{\extcalls}{\mbox{$\mathit{ExtCalls}$}}
\newcommand{\extorder}{\mbox{$\mathit{ExtOrder}$}}
\newcommand{\orderconstraint}{\mbox{$\mathit{OrderConstr}$}}

\newcommand{\stageinputname}{\Toolchain Input\xspace}
\newcommand{\stageimplname}{\Module Package\xspace}
\newcommand{\stageverifyname}{\Module Verification\xspace}
\newcommand{\stageoutputname}{\Toolchain  Output\xspace}

\newcommand{\circnum}[1]{%
\tikz[baseline=(char.base)]{
\node[draw, circle, inner sep=1pt, font=\bfseries\small] (char) {#1};
}}

\newcommand{\stageinputnotitle}{\circnum{1}}
\newcommand{\stageimplnotitle}{\circnum{2}}
\newcommand{\stageverifynotitle}{\circnum{3}}
\newcommand{\stageoutputnotitle}{\circnum{4}}

\newcommand{\stageinput}{\stageinputnotitle \ \stageinputname}
\newcommand{\stageimpl}{\stageimplnotitle \ \stageimplname}
\newcommand{\stageverify}{\stageverifynotitle \ \stageverifyname}
\newcommand{\stageoutput}{\stageoutputnotitle \ \stageoutputname}

\newcommand{\misrac}{MISRA-C\xspace}

\newcommand{\scania}{Scania\xspace}

\newcounter{cfrule}
\newcounter{dfrule}
\newcounter{ccrule}
\newcounter{princ}
\newcounter{task}

\renewcommand{\thecfrule}{CFR\arabic{cfrule}}
\renewcommand{\thedfrule}{DFR\arabic{dfrule}}

\renewcommand{\theprinc}{\arabic{princ}}
\renewcommand{\thetask}{T\arabic{task}}

\newcommand{\cfrlabel}[1]{\refstepcounter{cfrule}\label{#1}\thecfrule}
\newcommand{\dfrlabel}[1]{\refstepcounter{dfrule}\label{#1}\thedfrule}

\newcommand{\princlabel}[1]{\refstepcounter{princ}\label{#1}}
\newcommand{\newprinc}[1]{\princlabel{#1}\theprinc}
\newcommand{\tasklabel}[1]{\refstepcounter{task}\label{#1}}
\newcommand{\newtask}[1]{\tasklabel{#1}\thetask}

\lstdefinelanguage{ISContract}{
    basicstyle=\ttfamily,
    keywords=[1]{module, entry_functions, entry_order, external_calls, external_call_order},
    keywords=[2]{void, int, float, double, long},
    keywordstyle=[1]\color{magenta},
    keywordstyle=[2]\color{blue!80!black}
}
\lstdefinestyle{ISContractInline}{
    language=ISContract,
    keywordstyle=[1]\color{black},
    keywordstyle=[2]\color{black}
}
\lstdefinestyle{CStyle}{
    language=C,
    keywordstyle=\color{magenta},
    basicstyle=\scriptsize\ttfamily,
    commentstyle=\color{gray!80!black},
    backgroundcolor=\color{gray!10!white},
    numbers=left,
    numberstyle=\tiny\color{gray!70!white},
    numbersep=4pt
}
\lstdefinelanguage{ACSL}{%
	escapechar={},
	literate=,
	breaklines=false,
	morekeywords={
    int, void, float, double, long, static, extern, volatile, struct, union,
    assert,assigns,assumes,axiom,axiomatic,behavior,behaviors,
		boolean,breaks,complete,continues,data,decreases,disjoint,ensures,
		exit_behavior,ghost,global,inductive,invariant,lemma,logic,loop,
		model,relational,predicate,reads,requires,sizeof,strong,struct,terminates,
		type,union,variant,uchar,byte,typically,\\result,\\old,\\at,\\valid,\\valid_read,
		\\separated,\\nothing,Pre,\\exists,\\forall,\\sum,\\numof},
	alsoletter={\\}
 }

\lstdefinestyle{ACSLStyle}{
    style=CStyle,
    language=ACSL,
    mathescape,
    basicstyle=\ttfamily\footnotesize
}

\todostyle{blue}{
  color=blue!20
}

\begin{document}

\title{Contract Based Verification of Non-functional Requirements for Embedded Automotive C Code}
\titlerunning{Contract Based Verification of Non-functional Requirements}
%
\author{Jesper Amilon\inst{1}
\and
Merlijn Sevenhuijsen\inst{1,2}
\and
Mattias Nyberg\inst{2}
\and
Karl Palmskog\inst{1}
}
\authorrunning{J. Amilon et al.}
%
\institute{KTH Royal Institute of Technology, Sweden\\
\email{\{jamilon,palmskog\}@kth.se}\and
TRATON AB, Sweden\\
\email{\{merlijn.sevenhuijsen,mattias.nyberg\}@scania.com}}
\maketitle              
\begin{abstract}
Code contracts provide a robust way to specify functional requirements of safety-critical software in embedded systems. For example, the ANSI/ISO C Specification Language (\acsl) can be used to specify the functional behavior of C code that is then formally verified by the Frama-C framework's \wpplugin plugin. However, non-functional requirements, such as restrictions on control flow and data flow, are also important for embedded systems safety. Untrusted code developed by subcontractors, junior developers, or generated by large language models can be verified by \wpplugin but may nevertheless call unsafe functions or use uninitialized program variables.
To address this problem, we constructed a set of general rules concerning requirements on control flow and data flow of C code in safety-critical embedded systems. Our rules are orthogonal to popular C rulesets such as \misrac and center on modules and their interactions through interfaces. To enable rule checking, we propose an interface specification contract language for C modules. 
We implemented a \framac plugin which takes as input a C module and its interface specification contract and checks our rules, ensuring, e.g., that only functions permitted by the contract are called by the module.
We integrated our checker into a \toolchain to enable specification and verification of module contracts and \acsl contracts for untrusted code. 
We report on two case studies on safety-critical C code using software from \scania trucks, where we defined module contracts and \acsl function contracts based on informal system requirements and verified them using our \toolchain. 

\keywords{C  \and ACSL \and Control flow \and Data flow \and Formal verification \and Non-functional requirements \and Frama-C \and Code contracts}
\end{abstract}

\section{Introduction}
\label{sec:introduction}

Functional requirements of a program or system intuitively express what a system is supposed to \emph{do}, such as returning some output for a particular input.
In contrast, \emph{non-functional} requirements intuitively express how a system or program must be constituted. Such requirements may concern, e.g., program control flow and data flow restrictions, bounds on execution time~\cite{Abella2015} and other resources such as memory~\cite{Carbonneaux2014}, or even details on formatting of source code.

Functional requirements for safety-critical software, such as for embedded C code in automotive systems, can be expressed using \emph{code contracts} and then formally verified for adherence to these contracts~\cite{Lidstrom2023,Ung2024}. For example, contracts for functions in C code can be written in the ANSI/ISO C Specification Language (\acsl)~\cite{acsl_reference} and verified using the \framac framework~\cite{Baudin2021}, e.g., using the \wpplugin plugin for deductive verification~\cite{Kirchner2015}.
However, due to their diverse nature, non-functional requirements are difficult to capture in a single contract language and framework. For example, information flow security may be expressed at code level using type annotations, while execution time bounds may be expressed at system level using scheduler assumptions. In industrial practice, non-functional requirements are therefore often enforced by manual code reviews and tools like linters and other static checkers, such as \cppchecker~\cite{cppcheck}.

Non-functional requirements, in particular on control flow and data flow, are important in safety-critical automotive C code and explicitly mentioned by standards such as ISO~26262~\cite{iso26262}. 
Requirements on control flow in embedded systems software frequently concern inter-module properties, e.g., that a software module controlling some part of an Electronic Control Unit (ECU) does not inadvertently call certain functions in other modules. Formal verification of \acsl contracts does not rule out such undesirable calls, since the effect of the call may be independent of the properties that the contract requires.

Explicit specification and checking of such essential non-functional requirements may not be necessary when embedded software is developed by experienced engineers. Code reviews may even go beyond what is necessary and address maintainability and other less safety-critical issues. But when development includes untrusted code, e.g., as developed by subcontractors or junior engineers, or generated using large language models (LLMs), explicit specification and verification become necessary to ensure system-level properties~\cite{Kambhampati2024}. 

In this paper, based on our analysis of deployed embedded systems software written in C for ECUs in \scania trucks, we construct a set of general rules for non-functional requirements of C code, centered on \emph{modules} defined as pairs of \hfiles (declaring function signatures and datatypes) and \cfiles (containing data and function definitions). The rules are distilled from general requirements governing \scania embedded C code and internal coding standards, and capture properties that were found to be violated when generating code using LLMs~\cite{Sevenhuijsen2025b}. They are orthogonal to rulesets such as the Power of 10~\cite{Holzmann2006} and MISRA-C~\cite{MISRA2004}. 

To enable rule checking, we define a \emph{module Interface Specification (IS) contract language} that bridges the gap between C functions considered separately (as in ACSL) and modules of C code in embedded systems. Conventionally, such a module realizes control software for a single ECU and externally exposes a subset of its C functions to other modules. We introduce IS \emph{contracts}, which constitute one step towards bridging individual \acsl function contracts with automotive system (vehicle) level contracts.
Furthermore, we present a \framac plugin dubbed \plugin that checks our rules for a C module using an \iscontract. The plugin is released as open source software~\cite{amilon2026vernfr}.

We integrated \plugin into a \toolchain to enable practical verification of both module \iscontracts and \acsl function contracts. The \toolchain is based on using a collection of \emph{critics} on \emph{untrusted} C code~\cite{Kambhampati2024,Sevenhuijsen2025b} to ensure that the code satisfies properties from \emph{trusted} specifications. 
We evaluate our \toolchain through two case studies on modules in ECU controller software from \scania trucks. For each module, we annotate the entry-point functions with \acsl contracts and write module \iscontracts based on informal requirements. We then check the latest available \scania module implementation code using \plugin, \framac \wpplugin,  and \cppchecker. Finally, we perform preliminary experiments on generating case study module code from specifications using two popular LLMs, resulting in implementations that satisfy all verification checks.

In summary, we make the following contributions:
\begin{enumerate}
    \item A ruleset capturing key non-functional requirements of embedded C code related to control flow and data flow,
    \item a contract language for expressing rule-based properties of interfaces and implementations of C modules used in embedded systems,
    \item a novel contract-based checker dubbed \plugin for our rules, which we integrated into a \toolchain for verification of untrusted C code, 
    \item an evaluation of our contract-based specification and verification approach and \toolchain through two case studies on real-world embedded application modules from automotive ECUs in Scania trucks.
\end{enumerate}

\section{Background}
\label{sec:background}
In this section, we provide some background on automotive embedded C code and its specification and verification using ACSL, \framac, and other tools.

\subsection{Embedded control software in \scania trucks}
\label{sec:automotive_software}

\begin{figure}[ht]
    \includegraphics[width=\textwidth]{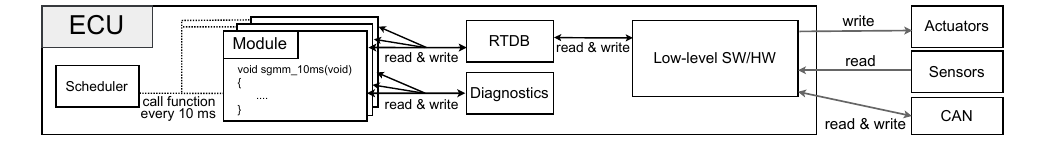}
    \caption{ECU architecture with internal and external interactions.}
    \label{fig:ecu}
\end{figure}

\scania trucks contain tens of Electronic Control Units (ECUs) which run software, called \emph{application modules}, that control and interact with physical actuators and sensors.
Each ECU has a static scheduler that repeatedly calls an \emph{entry-point function} of each application module every 10 ms. Application modules inside an ECU frequently interact with each other and with infrastructure software, primarily using a real-time database (RTDB) module and a diagnostics module. Typically, an application module invoked by the scheduler runs an entry-point function that reads values from RTDB, performs some computations, and writes back the results to RTDB.
Inter-ECU interaction is based on signals using the CAN protocol~\cite{ISO:CAN}.
\autoref{fig:ecu} illustrates the embedded systems architecture used in \scania trucks for ECUs and their application modules.

An application module implementation is typically written in the C language and contains code that has no unbounded loops or recursion and is executed sequentially. The subset of C used for application modules in \scania trucks roughly corresponds to \misrac (described in \autoref{sec:nfr}), but with additional conventions and rules.

\subsection{Modules and C code}
\label{sec:module}
The embedded systems concept of an application module used in \scania trucks does not precisely map to any definition in a language standard, such as the C99 standard that we target~\cite{ISO:C99}.
C99 instead divides software
into \emph{translation units}. Roughly speaking, a translation unit consists of a \cfile and, transitively, all files included with the \lstinline!#include!-directive. 
In practice, C software is often organized into implementation files (\cfiles) and associated header files (\hfiles). In this approach, a \cfile contains definitions, while a corresponding \hfile contains the declarations and macro language definitions intended to be shared with other translation units \cite{gcc-header-files}. The declarations of other translation units are imported by including the corresponding \hfiles. We adhere to this convention, and operationalize an automotive application module as a \emph{pair} of a \cfile and an \hfile of C code. Such a pair is then also a C99 translation unit.

When declaring functions and variables in C, \emph{storage specifiers} can be used to decide visibility to other translation units. Specifically, global declarations using the \lstinline!static! storage specifier are not visible
to other translation units. Global declarations without static storage specifier are visible to other translation units. From the module perspective, functions declared without the \lstinline!static! storage specifier can be viewed as entry points to the module and functions with \lstinline!static! storage specifier as \emph{local} functions.

\subsection{\acsl and \framac}
\label{sec:acsl_framac}
\noindent The ANSI-C Specification Language (\acsl)~\cite{acsl_reference} allows specifying properties of C~code. ACSL specifications are written in source code files as comments starting with \lstinline!@!. ACSL is based on first-order logic and includes function contracts, loop invariants, assertions, predicates, and logical functions. \autoref{fig:acsl_contract_rtdb_saturate} shows an example ACSL contract for the function \lstinline$update_sat$. The function returns either an incremented or a decremented value of the variable \lstinline!timer!, depending on the condition \lstinline!cond!. If the new value lies outside the interval \lstinline$[min, max]$, the result is truncated to the corresponding extreme point. The function is specified with an ACSL contract where the \emph{requires clause} states that the two extreme points form a non-empty interval contained in \lstinline$(INT_MIN, INT_MAX)$, the \emph{assigns clause} states that no global memory locations can be modified, and the \emph{ensures clause} specifies how the output should relate to the inputs.
\begin{figure}[!ht]
    \centering
\begin{lstlisting}[style=ACSLStyle, basicstyle=\scriptsize\ttfamily,numbers=none]
/*@ requires INT_MIN < min < max && min <= timer <= max < INT_MAX;
    assigns \nothing;
    ensures 
      (cond == 0 ==> \result == \max(min, \old(timer) - 1)) &&
      (cond != 0 ==> \result == \min(max, \old(timer) + 1)); */
int update_sat(int cond, int timer, int min, int max) {
  int res = timer;
  if (cond) {res += 1;} else {res -= 1;}
  if (res > max) res = max;
  if (res < min) res = min; 
  return res;
}
\end{lstlisting}
    \caption{Example C function annotated with an \acsl contract.}
    \label{fig:acsl_contract_rtdb_saturate}
\end{figure}

\framac is a platform for analysis and verification of C code and ACSL specifications~\cite{Baudin2021}. \framac is comprised of a kernel and numerous plugins. The kernel is responsible for parsing the C sources and the ACSL specifications into a combined Abstract Syntax Tree (AST), which can be analyzed by the plugins. 
Before building the AST, \framac runs the preprocessor from the compiler GCC~\cite{gccmanual} on the C sources, e.g., to resolve compiler directives such as \lstinline!#define!. 

The \framac \wpplugin plugin~\cite{wp_manual} is used to verify that a function satisfies its ACSL function contract. 
\wpplugin is based on a weakest-precondition calculus and verifies a program by generating a set of proof obligations, which must be discharged for the verification to succeed. Verification with \wpplugin is typically combined with checking absence of run-time errors, such as array-out-of-bounds or overflows. The command line option \lstinline!wp-rte! instructs \wpplugin to generate proof obligations also for such run-time errors. To discharge the proof obligations, \wpplugin relies on SMT solvers, such as Alt-Ergo~\cite{conchon_alt-ergo_2018} and Z3~\cite{z3}.
The \framac \eva plugin~\cite{eva_manual}, based on abstract interpretation, is used to calculate domain ranges for variables. The result from \eva is often taken as input by other plugins, e.g., to show the absence of runtime errors.

\acsl~Importer is a plugin for \framac which allows \acsl specifications separately from C code in \texttt{.acsl} files~\cite{acslimporter}. The plugin enables listing all function contracts for modules, but does not enable expressing module properties.

\subsection{Non-functional requirements and C static checking}
\label{sec:nfr}
For safety-critical C modules in ECUs, we consider \emph{functional correctness} as a baseline, which we define as adhering to given requirements on functional behavior. For example, functional requirements may be defined in natural language and then translated to ACSL contracts on module functions~\cite{Ung2024}.
In contrast, non-functional requirements on C modules in ECUs come from a variety of sources and are infeasible to express at code level using a single specification language. 

Several standards provide guidance on desirable non-functional properties of C code. 
ISO 26262~\cite{iso26262} is a standard for safety in automotive systems and includes general requirements on software, but these requirements must be contextualized significantly to be applied at the code level.
The \misrac standard~\cite{MISRA2004} defines, through a set of \emph{guidelines}, a subset of C that reduces the risk for undesirable program behavior and errors. 
\misrac rules distinguish between \emph{directives} and \emph{rules}. Directives are guidelines that may require analysis beyond the source code, such as documenting how the code relates to external requirements. Rules are intended to be checkable on the code itself, and typically aim to restrict usage of C constructs that are complex and error-prone. A subset of the \misrac rules can be checked using static analysis tools, such as \cppchecker~\cite{cppcheck}. Each guideline is further classified as \emph{mandatory}, \emph{required}, or \emph{advisory}. Mandatory guidelines must be followed without exception to conform to \misrac, while required guidelines are expected to be followed but may be deviated from with justification, and advisory guidelines represent recommended practice. \cppchecker and other \misrac checkers commonly report the classification of each violation found in the checked code.

\begin{table}[!htb]
    \caption{Principles for interfaces in C code used in embedded systems.}
        \label{tab:principles}
\scriptsize
    \centering
    \begin{tabular}{m{0.04\textwidth}m{0.38\textwidth}@{\hspace{0.02\textwidth}}m{0.54\textwidth}}
      \toprule
       & \textbf{Principle}  & \textbf{Motivation} \\
      \toprule
      \newprinc{prn:interaction_fun_only} &
      All interaction with other modules shall be manifested through function calls. &
      Interaction using function calls, rather than reading and writing to global variables, simplifies modular code development.
      \\ \midrule
      \newprinc{prn:standardized} &
      The interface of the module shall be standardized. &
      Standardization increases code readability and disallows, e.g., unconventional usage of \hfiles and \cfiles, and ensures that data flow between modules does not cause unexpected behavior.
      \\ \midrule
      \newprinc{prn:entry_points} &
      The \interfacespec shall define the entry-point functions for the module, and exactly these shall be callable from other modules. &
      This ensures that the interface is clear and well-defined.
      \\ \midrule
      \newprinc{prn:permitted_calls} &
      The \interfacespec shall define exactly the functions in other modules that are permitted to be called; and the module shall adhere to this. &
    First, a function signature may be defined in several \hfiles, so the \interfacespec shall indicate which one to use. 
    Second, it may be necessary to permit only a subset of the functions in an \hfile to be called.
      \\ \midrule
    \newprinc{prn:call_order} &
      The \interfacespec shall constrain the call order of function calls to other modules, and the module shall adhere to this. &
      The behavior of other modules may depend on the call order, for example, to ensure that variables are initialized before they are read. 
    \\ \midrule
    \newprinc{prn:standalone} &
    The \interfacespec shall constitute a single standalone artifact.  &
    This simplifies, strengthens, and clarifies the concept of an \interfacespec. 
    \\
    \bottomrule
    \end{tabular}
\end{table}
\section{Interfaces: from Principles to Rules}
\label{sec:rules}
In this section, we develop \emph{principles} and \emph{rules} for \emph{module interfaces} in embedded C, which constrain how modules interact with each other. The principles capture non-functional properties of modules related to control flow and data flow, and are meant to complement requirements for a module's functional behavior, e.g., as expressed in \acsl function contracts. We aim for the principles to enable \emph{modular} development and verification of modules~\cite{deRoever1998}. To this end, we require modules to be equipped with an Interface Specification (IS). We first introduce the principles, and then distill them into \emph{coding rules} and a \emph{contract language} for interface specifications, where the latter is defined in \autoref{sec:module-contracts}.

\subsection{Interface principles}
Our principles for module interfaces and module \interfacespecs are given in \autoref{tab:principles}. In accordance with \autoref{sec:module}, we adopt the convention that a module consists of a pair of a \cfile and an \hfile. 

The principles are largely based on experience from verifying automotive embedded software~\cite{Ung2024,Lidstrom2017,Amilon2025}. We also consider the long-standing convention of employing \emph{information hiding} in modular software development~\cite{parnas1983}. Information hiding means that implementation details that may be subject to change should be hidden. For example, in \autoref{prn:interaction_fun_only}, we treat global variables within a module as implementation details that can be changed at any time, whereas the entry-point functions should provide a stable interface across module versions.

\begin{table}[tb]
\scriptsize
\centering
\caption{Rules for control flow.}
\label{tab:cf_rules}

\begin{tabular}{m{0.07\textwidth}@{\hspace{8pt}}m{0.62\textwidth}@{\hspace{8pt}}m{0.25\textwidth}}%
\toprule
\textbf{ID} & \textbf{Rule} & \textbf{Rationale} 
\\
\toprule
\cfrlabel{cfr:ext_calls} &
The module shall only call external functions permitted by the \interfacespec. &
Ensures \autoref{prn:permitted_calls}. \\
\midrule
\cfrlabel{cfr:ext_calls_order} &
The order of external calls shall respect the order constraints in the \interfacespec, given an assumed call order of the entry-point functions of the module. &
Ensures \autoref{prn:call_order}. 
\\
\midrule
\cfrlabel{cfr:fnc_ptrs}
&
Function pointers shall not be used.
&
Prerequisite for checking \ref{cfr:ext_calls} and \ref{cfr:ext_calls_order}.
\\
\midrule
\cfrlabel{cfr:fnc_defs} &
No function definitions or initializations in the \hfile. 
& 
Helps ensure \autoref{prn:standardized}.
\\
\midrule
\cfrlabel{cfr:only_h_files}
&
Include only \hfiles. & Helps ensure \autoref{prn:standardized}.
\\
\midrule
\cfrlabel{cfr:entry_points_declared} &
Entry-point functions shall be declared in the \hfile without the static storage-class specifier. & Helps ensure \autoref{prn:entry_points}.
\\
\midrule
\cfrlabel{cfr:entry_points_defined} &
Entry-point functions shall be defined in the \cfile. & Helps ensure  \autoref{prn:entry_points}.
\\
\midrule
\cfrlabel{cfr:non_entry_static} &
Non-entry-point functions shall be declared in the \cfile with a static storage class specifier. & Helps ensure  \autoref{prn:entry_points}.
\\
\midrule
\cfrlabel{cfr:only_entry_points_hfile} &
Only entry-point functions in the \interfacespec shall be declared in the \hfile.
 & Helps ensure  \autoref{prn:entry_points}.
\\
\midrule
\cfrlabel{cfr:extern_keyword} &
The keyword \lstinline!extern! shall not be used, except for the declaration of entry-point functions. & Helps ensure \autoref{prn:standardized}.
\\
\midrule
\cfrlabel{cfr:entry_point_type} &
The types for the function declarations in the \hfile and the \interfacespec shall match exactly. & Helps ensure \autoref{prn:entry_points}.
\\
\bottomrule
\end{tabular}
\end{table}

\begin{table}[!htb]
\scriptsize
\centering
\caption{Rules for data flow.}
\label{tab:df_rules}

\begin{tabular}{m{0.07\textwidth}@{\hspace{8pt}}m{0.38\textwidth}@{\hspace{8pt}}m{0.48\textwidth}}
\toprule
\textbf{ID} & \textbf{Rule} & \textbf{Rationale} 
\\
\toprule
\dfrlabel{dfr:variables_static} &
All variables with file-global scope shall be declared in the \cfile with the static storage specifier. & 
Implies that module interaction through global variables is prohibited. Helps ensure \autoref{prn:interaction_fun_only}.
\\
\midrule
\dfrlabel{dfr:ptr_arithmetics} &
Pointer arithmetic shall not be used. 
&
Necessary to ensure that all interactions with other modules are through function calls. Helps ensure \autoref{prn:interaction_fun_only}. 
\\
\midrule
\dfrlabel{dfr:type_casting_ptrs}
& Explicit type casting from pointers and to pointers shall not be used. & Necessary to ensure that all interaction with other modules is through function calls. Helps ensure \autoref{prn:interaction_fun_only}. 
\\
\midrule
\dfrlabel{dfr:variables_initialised} &
A static declared variable shall not be read before it is explicitly initialized or written to.  
& All variables shall be initialized to a meaningful value; we do not rely on default initialization to 0. Helps ensure \autoref{prn:standardized}.
\\
\midrule
\dfrlabel{dfr:ptr_literals} &
Pointer literals shall not be used. & Necessary to ensure that all interaction with other modules is through function calls. Helps ensure \autoref{prn:interaction_fun_only}. 
\\
\midrule
\dfrlabel{dfr:use_typedefs} 
& 
Typedefs (type aliases) shall be used whenever applicable.
&
Ensures that the module interacts in a predictive way (\autoref{prn:standardized}). For example, using typedefs can ensure that the same module behaves as expected across platforms with varying bit sizes. 
\\
\bottomrule
\end{tabular}
\end{table}

\subsection{Interface rules}
We operationalize the principles from \autoref{tab:principles} into 17 rules that are necessary to follow for a module to adhere to the 6 principles. We split the rules into two categories: in \autoref{tab:cf_rules}, rules that constrain the control flow between modules, and in \autoref{tab:df_rules}, the rules that constrain the data flow between modules. For each rule, we also indicate how they relate to the 6 principles. 

Given an \interfacespec and a module, our intention is that these rules can be checked using conventional methods for static analysis or formal verification. For the kind of embedded software that we target, we also expect that the checking of these rules will be decidable. For general C code, the checking of these rules will be undecidable for programs with, e.g., recursion and function pointers. In \autoref{sec:implementation}, we present a \framac plugin that verifies a subset of our rules.
The list of rules is, intentionally, not complete with respect to adherence to the principles. Specifically, we have not included rules that are already covered by the \misrac standard~\cite{MISRA2004}. For example, avoiding dynamic memory allocation, i.e., \texttt{malloc} and \texttt{calloc}, could help maintain \autoref{prn:interaction_fun_only}, but this is already covered by \misrac directive 4.12. Similarly, array out-of-bounds accesses may corrupt the interface of a module, but avoiding array out-of-bounds accesses is part of the general \misrac directive 4.1: \emph{Minimize run-time errors}.

\section{Module Interface Specification Contract Language}
\label{sec:module-contracts}
Based on the principles in \autoref{sec:rules}, we propose in this section a language for interface specification (IS) contracts.
We choose to define the interface specifications as contracts since they need to capture both \emph{assumptions} and \emph{guarantees}. For example, in \autoref{prn:call_order}, we want to \emph{assume} the call order for the entry-point functions, and \emph{guarantee} the call order of the external calls. 

The intended use case of our \iscontracts is to provide the necessary context for verification of the rules \ref{cfr:ext_calls}, \ref{cfr:ext_calls_order},  \ref{cfr:entry_points_declared}, \ref{cfr:entry_points_defined}, and \ref{cfr:entry_point_type} from \autoref{sec:rules}. The remaining rules can be verified by analyzing the source code without module-related context.

\begin{figure}[!htb]
    \centering
\begin{lstlisting}[language=ISContract, basicstyle=\scriptsize\ttfamily, backgroundcolor = \color{gray!10!white}]
module tmon {
  entry_functions: { void tmon_init(void), int tmon_step(void) }
  entry_order: { tmon_init < tmon_step }
  external_calls: {
    sensors.h: { void tmon_sens_create(void), int tmon_sens_read(void) },
    warnings.h: { void tmon_warn_create(void), void tmon_warn_write(int) },
    utils.h: { int update_sat(int, int, int, int) }
  }
  external_call_order: { 
    tmon_sens_create < tmon_sens_read, tmon_warn_create < tmon_warn_write
  }
}
\end{lstlisting}
    \caption{Module Interface Specification (IS) contract \isfile for a timed monitor.}
    \label{fig:mod_interface_contract}
\end{figure}

\begin{figure}[!htb]
    \centering
\begin{subfigure}[t]{0.45\linewidth}
\begin{lstlisting}[title=\texttt{tmon\_a.h},style=CStyle,linewidth=\linewidth,showlines=true]
void tmon_init(void);

int tmon_step(void);
\end{lstlisting}
\vspace{-1em}
\begin{lstlisting}[title=\texttt{tmon\_a.c},style=CStyle,linewidth=\linewidth,showlines=true]
#include "sensors.h"
#include "warnings.h"
#include "utils.h"

static int timer;

void tmon_init(void){
  timer = 5; 
  tmon_sens_create();
  tmon_warn_create();
}

void tmon_step(void){
  int s = tmon_sens_read();
  timer = update_sat(s,timer,0,10);
  if (timer == 10) { 
    // activate warning
    tmon_warn_write(1);
  } else if (timer == 0) { 
    // deactivate warning
    tmon_warn_write(0);
  } else {
    // do nothing
  }
}
\end{lstlisting}
    \caption{Implementation that satisfies the \texttt{tmon} \iscontract.}
    \label{fig:mod_good_example}
\end{subfigure}
\hfill
\begin{subfigure}[t]{0.45\linewidth}    
\begin{lstlisting}[title=\texttt{tmon\_b.h},style=CStyle,breaklines=true]
void tmon_init(void); 
int get_tmon_timer(void);
static int tmon_step(void);
\end{lstlisting}
\vspace{-1em}
\begin{lstlisting}[title=\texttt{tmon\_b.c},style=CStyle,breaklines=true]
#include "sensors.h"
#include "warnings.h"
#include "utils.h"

int timer;

void tmon_init(void){
  tmon_sens_create();
}

static void tmon_step(void){
  int s = tmon_sens_read();
  timer = update_sat(s,timer,0,10);
  if (timer == 10) { 
    // activate warning
    tmon_warn_write(1);
  } else if (timer == 0) { 
    // deactivate warning
    xmon_warn_write(0);
  }
}

int get_tmon_timer(void) {
  return timer;
}
\end{lstlisting}
    \caption{Implementation that does not satisfy the \texttt{tmon} \iscontract.}
    \label{fig:mod_bad_example}
\end{subfigure}
    \caption{Implementations of the timed monitor (\texttt{tmon}) module from \autoref{fig:mod_interface_contract}.}
    \label{fig:mod_example}
\end{figure}

\subsection{Running example}
The \iscontract in \autoref{fig:mod_interface_contract} for the module \lstinline!tmon! serves as a running example together with two implementations: \autoref{fig:mod_good_example}, which satisfies the \iscontract, and \autoref{fig:mod_bad_example}, which violates the \iscontract. The \iscontract specifies two entry-point functions, \lstinline$tmon_step$ and \lstinline$tmon_init$.  
The \lstinline$tmon_init$ function initializes the sensor and warning functionality, and is intended to be called at startup. \lstinline$tmon_step$ is intended to be called repeatedly by a scheduler, and implements a simple timed monitor. The \lstinline$tmon_step$ function reads the value of a sensor and, if the value is non-zero, the timer is incremented; otherwise, it is decremented.  If the timer reaches 10, a warning is activated; if the timer reaches 0, the warning is deactivated. The purpose of the timer is to act as a \emph{delay filter}, ensuring that the warning is activated and deactivated only after several consecutive consistent readings. Such a delay filter is common in automotive software to avoid false warnings or flickering behavior. Indeed, \lstinline$tmon$ is a simplified version of our real-world case study module SFLD, presented in \autoref{sec:case_studies}.

The external calls specified in the \iscontract allow the module to communicate with the sensor, activate the warning signal, and use the utility function \lstinline$update_sat$ from \autoref{fig:acsl_contract_rtdb_saturate}, whose declaration is in the header file \texttt{utils.h}.

\begin{figure}[tb]
    \centering
\begin{equation*}
    \begin{array}{rll}
     \nonterminal{\mathit{ISContract}} &::=&   \texttt{module}\; \nonterminal{\mathit{Id}} \; \texttt{\{}\,
      \nonterminal{\entrypoints} 
      \nonterminal{\entryorder}
      \nonterminal{\extcalls}   \nonterminal{\extorder}\,\texttt{\}} \\
    \nonterminal{\entrypoints} & ::=& \texttt{entry\_functions}\,\texttt{:}\; \texttt{\{}\, \nonterminal{\mathit{FunDecl}}^\star \,\texttt{\}} \\
    \nonterminal{\entryorder} & ::=& \texttt{entry\_order}\,\texttt{:} \;\texttt{\{}\,{\nonterminal{\orderconstraint}^\star}\,\texttt{\}} \\ 
    \nonterminal{\extcalls} & ::=& \texttt{external\_calls}\,\texttt{:}\; \texttt{\{}\,\nonterminal{\extmodule}^\star\,\texttt{\}} \\
    \nonterminal{\extmodule} & ::=& \nonterminal{\mathit{HeaderName}}\texttt{:}\, \texttt{\{}\,\nonterminal{\mathit{FunDecl}}^\star\,\texttt{\}} \\
    \nonterminal{\extorder} & ::=& \texttt{external\_call\_order}\,\texttt{:}\; \texttt{\{}\,\nonterminal{\orderconstraint}^\star\,\texttt{\}}\\ 
\nonterminal{\orderconstraint} & ::=& \nonterminal{\mathit{Id}} < \nonterminal{\mathit{Id}} \mid 
    \nonterminal{\mathit{Id}} > \nonterminal{\mathit{Id}} \\
    \end{array}
\end{equation*}
    \caption{Grammar for \iscontracts, where the $\star$ operator means a possibly empty comma-separated list of what came before it. 
    }
    \label{fig:syntax_interface_contracts}
\end{figure}

\subsection{Contract language syntax and semantics}
We define the abstract syntax of \iscontracts by the grammar shown in \autoref{fig:syntax_interface_contracts}. We assume the nonterminals $\nonterminal{\mathit{FunDecl}}$ and $\nonterminal{\mathit{Id}}$ are defined according to the C99 standard for function declarations and identifiers, respectively. The nonterminal $\nonterminal{\mathit{HeaderName}}$ is assumed to be a header file name with file ending \texttt{.h} that conforms to \misrac requirements.
We provide an informal semantics for our contract language, and leave an explicit formal semantics as future work. 
Given an \iscontract $T$, we describe its semantics by considering when (and when not) a module $m$ satisfies the contract. 
We say that $m$ satisfies $T$ precisely when the following three conditions hold:
\begin{enumerate}[label=(\roman*)]
    \item Precisely the entry-point functions in $T$ are declared in the \hfile and defined in the \cfile of $m$. 
    \item All external functions invoked by the \cfile in $m$ are permitted by $T$.
    \item Assuming that invocations of the entry-point functions specified in $T$ follow the entry order defined in $T$, then the invocations of the external functions by $m_c$ respect the external call order in $T$.
\end{enumerate}
For the contract in \autoref{fig:mod_interface_contract}, we see that the module implementation in \autoref{fig:mod_good_example} respects (i)-(iii), and hence satisfies the contract.
The module implementation in \autoref{fig:mod_bad_example}, on the other hand, violates all conditions (i)-(iii). Consequently, it does not satisfy the contract.

Regarding (i), we consider a function to be declared if there is a function declaration in the \hfile of $m$ and defined in the \cfile without a static storage specifier. In \autoref{fig:mod_bad_example}, we see that the implementation violates (i) in two places. First, it declares and defines the function \lstinline$mod_error_status$, which is not mentioned in the \iscontract. Second, the entry-point \lstinline$tmon_step$ is declared with the static keyword. 
Regarding (ii), \autoref{fig:mod_bad_example} violates this due to the illicit invocation of \lstinline$xmon_warn_write$, where we assume \lstinline$xmon_warn$ to be some other RTDB variable.
Regarding (iii), \autoref{fig:mod_bad_example} violates this, since the \lstinline$tmon_init$ function fails to call \lstinline$tmon_warn_create$. Thus, when \lstinline$tmon_step$ is called, the constraint \lstinline$tmon_warn_create < tmon_warn_write$ is violated. 

We can also relate conditions (i)-(iii) to the rules in \autoref{sec:rules}. Condition~(i) encompasses \ref{cfr:entry_points_declared}, \ref{cfr:entry_points_defined}, and \ref{cfr:entry_point_type},  condition~(ii) encompasses \ref{cfr:ext_calls}, and condition~(iii) encompasses \ref{cfr:ext_calls_order}. 

\section{\plugin: A \framac Plugin for Non-Functional Verification}
\label{sec:implementation}
\begin{table}[tb]
\scriptsize
    \centering
     \renewcommand{\arraystretch}{1.2}
\caption{Implemented verification tasks in \plugin. The column ``\isfile'' indicates if the task depends on an IS contract file.}
\label{tab:verification_tasks}
    \begin{tabular}{m{0.6cm}m{8.52cm}@{\hspace{8pt}}m{1.60cm}c}
    \toprule
     & \textbf{Task} & \textbf{Rule} & \textbf{\isfile} \\
    \toprule
    \newtask{tsk:external_calls} & External function calls are permitted by the \isfile & \ref{cfr:ext_calls} & \checkmark
    \\
    \midrule
    \newtask{tsk:fun_ptrs} & Absence of function pointers. & \ref{cfr:fnc_ptrs} & --
    \\
    \midrule
    \newtask{tsk:no_fun_defs} & No function definitions in the \hfile. & \ref{cfr:fnc_defs} & --
    \\
    \midrule
    \newtask{tsk:only_hfiles} & Only \hfiles are included. & \ref{cfr:only_h_files} & --
    \\
    \midrule
    \newtask{tsk:all_entry_declared} & All entry points in the \isfile are declared. & \ref{cfr:entry_points_declared} & \checkmark 
    \\
    \midrule
    \newtask{tsk:all_entry_defined} & All entry points in the \isfile are defined.  & \ref{cfr:entry_points_defined} & \checkmark 
    \\
    \midrule
    \newtask{tsk:non_entry_static} & Non-entry points are declared in the \cfile with static storage specifier. & \ref{cfr:non_entry_static}, \ref{cfr:only_entry_points_hfile} & \checkmark 
    \\
    \midrule
    \newtask{tsk:vars_static} & Variables have static storage specifier. & \ref{dfr:variables_static} & --
    \\
    \midrule
    \newtask{tsk:proper_init} & All memory locations are explicitly initialized or written to before they are read. & \ref{dfr:variables_initialised} & \checkmark
    \\
    \midrule
    \newtask{tsk:ptr_literals} & Absence of pointer literals. & \ref{dfr:ptr_literals} & --
    \\
    \midrule
    \newtask{tsk:typedefs} & Typedefs are always used when possible. & \ref{dfr:use_typedefs} & --
    \\
    \bottomrule
\end{tabular}
\end{table}
We now present  \plugin~\cite{amilon2026vernfr}, a 
 \framac plugin written in the OCaml programming language~\cite{leroy2018ocaml}, which implements a series of custom analyses tailored for verification of the rules in \autoref{sec:rules}. 
 In its current state, \plugin implements 11 verification tasks, listed in \autoref{tab:verification_tasks}. The table also indicates which of the rules in \autoref{sec:rules} the task targets, and whether the task requires an \isfile. Each task can be run as a command-line option when running Frama-C. We also provide a script, which takes as input a \cfile, an \hfile, and an \isfile, and runs all verification tasks.

The verification tasks are, in general, implemented by traversing the AST built by the \framac kernel.
For Task~\ref{tsk:proper_init}, we utilize the \framac plugin \eva, which can compute the memory locations, including global variables, that are inputs and outputs for each function. For each function and each input global variable, we then check that it is either explicitly initialized in the \cfile or that it is an output of some predecessor function according to the order constraints in the \iscontract. 
Task~\ref{tsk:only_hfiles} is implemented with a regex search in the source files.

Currently, the tasks do not fully cover the rules in \autoref{sec:rules}. Rules
\ref{cfr:extern_keyword}, \ref{dfr:ptr_arithmetics}, and \ref{dfr:type_casting_ptrs} are difficult to implement in the Frama-C framework due to pre-processing implemented by the Frama-C kernel. For example, array accesses are converted into pointer arithmetic expressions; hence, verifying the absence of pointer arithmetic is difficult without false positives.
We have also yet to implement verification of \ref{cfr:ext_calls_order}, that the module respects the call order constraints in the \iscontract. We expect this to be the most difficult rule to implement since it requires symbolic analysis to infer the possible call order sequences in a program.

Returning to our example module in \autoref{fig:mod_example}, the implementation in \autoref{fig:mod_good_example} passes all the tasks T1-T11, while for the other implementation in \autoref{fig:mod_bad_example}, verification fails for tasks \ref{tsk:external_calls}, \ref{tsk:all_entry_declared}, \ref{tsk:all_entry_defined}, \ref{tsk:non_entry_static}, \ref{tsk:vars_static}, and \ref{tsk:proper_init}.

\section{\Toolchain for Module Development and Verification}
\label{sec:workflow}
To enable practical embedded systems module development and verification with \plugin and \framac, we propose a workflow supported by a \toolchain, which we describe in this section. 
Recall that in our setting, a \emph{\module} is defined as a pair of a \hfile and a \cfile, as described in \autoref{sec:module}. To reason about the development and verification process for such modules, we distinguish between \emph{trusted} and \emph{untrusted} artifacts. Requirements, specifications, and verification tools are considered trusted, while module implementations are considered untrusted and must be verified. These implementations are produced by an \emph{\agent}, which may be a human programmer or an automated system such as an LLM.

Verification is carried out by a collection of \emph{critics}, which are components that analyze code with or without explicit specifications. A module is considered \emph{verified} if all critics report success; otherwise, it is not verified. Failure to verify does not necessarily imply incorrectness, but rather that correctness could not be established. The implementation, together with the associated specifications and header files, is referred to as a \emph{module package}, which serves as the unit of verification.

\definecolor{OIblue}{RGB}{86,180,233}
\definecolor{OIyellow}{RGB}{240,228,66}
\definecolor{OIgray}{RGB}{120,120,120}
\tikzset{
trusted/.style={
    draw=OIblue!70!black,
    fill=OIblue!15,
    rounded corners=3pt,
    line width=0.5pt
},
inputitem/.style={
    draw=OIblue!70!black,
    fill=OIblue!15,
    line width=0.5pt,
    minimum width=30mm,
    minimum height=6mm,
    align=center,
    inner xsep=2mm,
},
codefiledraw/.style={
    draw=OIyellow!70!black,
    fill=OIyellow!25,
    line width=0.5pt
},
trustedfiledraw/.style={
    draw=OIblue!70!black,
    fill=OIblue!15,
    line width=0.5pt
},
untrustedcircle/.style={
    draw=OIyellow!70!black,
    fill=OIyellow!25,
    minimum size=20mm,
    circle,
    align=center
},
critic/.style={
    draw=OIblue!70!black,
    fill=OIblue!15,
    rounded corners=2.5pt,
    line width=0.5pt,
    minimum width=22mm,
    minimum height=6mm,
    align=center
},
packagebox/.style={
    draw=black!40,
    fill=black!2,
    rounded corners=3pt,
    line width=0.6pt
},
outputparser/.style={
    draw=OIblue!70!black,
    fill=OIblue!15,
    rounded corners=2.5pt,
    line width=0.5pt,
    minimum width=28mm,
    minimum height=8mm,
    align=center
},
logbox/.style={
    draw=OIblue!70!black,
    fill=OIblue!15,
    line width=0.5pt,
    minimum width=28mm,
    minimum height=14mm,
    align=left
},
verdictnonverified/.style={
    draw=OIgray!70!black,
    fill=OIgray!12,
    line width=0.5pt,
    minimum width=18mm,
    minimum height=18mm,
    align=center
},
verdictverified/.style={
    draw=black!70,
    fill=white,
    line width=0.5pt,
    minimum width=18mm,
    minimum height=18mm,
    align=center
},
criticgroup/.style={
    draw=black!40,
    fill=none,
    rounded corners=3pt,
    line width=0.6pt
},
stage/.style={
    draw=black!70,
    fill=white,
    circle,
    minimum size=6mm,
    inner sep=0pt,
    font=\small\bfseries
},
implgroup/.style={
    draw=black!40,
    dashed,
    rounded corners=3pt,
    line width=0.6pt,
    inner sep=3mm
}
}

\begin{figure}[!ht]
\centering
\resizebox{\textwidth}{!}{%
\begin{tikzpicture}[node distance=2.4mm]

\node[trusted, minimum width=32mm, minimum height=54mm] (input) {};

\node[font=\bfseries] at ([yshift=-3mm]input.north) {\stageinput};

\node (inputicon) at ([yshift=-8mm]input.north)
    {\includegraphics[width=0mm]{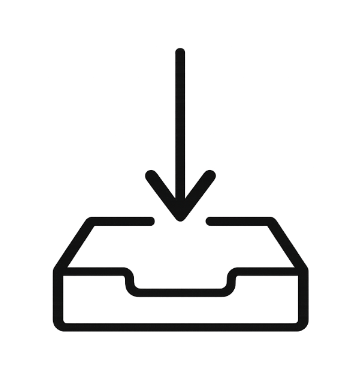}};

\node[inputitem, below=-2mm of inputicon] (iface) {Interface\\specification};
\node[inputitem, below=of iface] (nl) {Natural language\\specification};
\node[inputitem, below=of nl] (formal) {Formal\\specification};
\node[inputitem, below=of formal] (headers) {External\\ header files};

\node[untrustedcircle, below=5mm of input] (agent) {Agent\\
{\includegraphics[width=1cm]{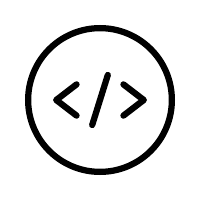}}
};

\draw[->, thick] (input.south) -- (agent.north);

\node[
    draw=OIyellow!70!black,
    fill=OIyellow!25,
    rounded corners=2pt,
    minimum width=5mm,
    minimum height=5mm,
    anchor=west
] (legend_untrusted_box) at ([xshift=-6mm,yshift=-28mm]agent.south west) {};

\node[anchor=west, right=2mm of legend_untrusted_box] (legend_untrusted_text) {Untrusted};

\node[
    draw=OIblue!70!black,
    fill=OIblue!15,
    rounded corners=2pt,
    minimum width=5mm,
    minimum height=5mm,
    anchor=west
] (legend_trusted_box) at ([xshift=39mm]legend_untrusted_box.west) {};

\node[anchor=west, right=2mm of legend_trusted_box] (legend_trusted_text) {Trusted};

\node[packagebox, minimum width=54mm, minimum height=90mm] (package)
    [right=12mm of agent, yshift=18mm] {};

\node at ([xshift=-22.5mm, yshift=-14.5mm]package.north) (modis) {};
\node[below=13mm of modis] (fspec) {};
\node[below=17mm of fspec] (hdrs) {};
\node[below=19mm of hdrs] (modc) {};
\node[below=13mm of modc] (modh) {};

\node[fill=black!10, fit= {(modc) ([yshift=-53mm]package.north) ([xshift=-2.5mm,yshift=3mm]package.south east)}] (modframe) {};

\node[font=\bfseries] (module) at ([yshift=-55mm]package.north) {\Module};

\begin{scope}[shift={(modis.center)}]
  \coordinate (modiswest) at (0,6mm);

  \draw[trustedfiledraw]
    (0,0) -- (0,12mm) -- (44mm,12mm) -- (47mm,9mm) -- (47mm,0) -- cycle;
  \draw[trustedfiledraw] (44mm,12mm) -- (44mm,9mm) -- (47mm,9mm);

  \node[anchor=west, font=\bfseries\large] at (1.2mm,8.8mm) {mod.is};
  \draw[OIblue!70!black] (1.2mm,6.5mm) -- (45.5mm,6.5mm);
  \node[anchor=north west, align=left, font=\ttfamily\footnotesize] at (1.2mm,5.5mm) {%
calls, order, ...};
\end{scope}

\begin{scope}[shift={(fspec.center)}]
  \coordinate (fspecwest) at (0,6mm);

  \draw[trustedfiledraw]
    (0,0) -- (0,12mm) -- (44mm,12mm) -- (47mm,9mm) -- (47mm,0) -- cycle;
  \draw[trustedfiledraw] (44mm,12mm) -- (44mm,9mm) -- (47mm,9mm);

  \node[anchor=west, font=\bfseries\large] at (1.2mm,8.8mm) {mod.acsl};
  \draw[OIblue!70!black] (1.2mm,6.5mm) -- (45.5mm,6.5mm);
  \node[anchor=north west, align=left, font=\ttfamily\footnotesize] at (1.2mm,5.5mm) {%
 requires ...   };
\end{scope}

\begin{scope}[shift={(hdrs.center)}]
  \coordinate (hdrswest) at (0,6mm);

  \draw[trustedfiledraw]
    (4mm,4mm) -- (4mm,16mm) -- (38mm,16mm) -- (41mm,13mm) -- (41mm,4mm) -- cycle;
  \draw[trustedfiledraw] (38mm,16mm) -- (38mm,13mm) -- (41mm,13mm);

  \draw[trustedfiledraw]
    (2mm,2mm) -- (2mm,14mm) -- (36mm,14mm) -- (39mm,11mm) -- (39mm,2mm) -- cycle;
  \draw[trustedfiledraw] (36mm,14mm) -- (36mm,11mm) -- (39mm,11mm);

  \draw[trustedfiledraw]
    (0,0) -- (0,12mm) -- (34mm,12mm) -- (37mm,9mm) -- (37mm,0) -- cycle;
  \draw[trustedfiledraw] (34mm,12mm) -- (34mm,9mm) -- (37mm,9mm);

  \node[anchor=west, font=\bfseries\large] at (1.2mm,8.8mm) {headers};
  \draw[OIblue!70!black] (1.2mm,6.5mm) -- (35.5mm,6.5mm);
  \node[anchor=north west, align=left, font=\ttfamily\footnotesize] at (1.2mm,5.5mm) {%
types.h ...};
\end{scope}

\begin{scope}[shift={(modc.center)}]
  \coordinate (modcwest) at (0,6mm);

  \draw[codefiledraw]
    (0,0) -- (0,12mm) -- (44mm,12mm) -- (47mm,9mm) -- (47mm,0) -- cycle;
  \draw[codefiledraw] (44mm,12mm) -- (44mm,9mm) -- (47mm,9mm);

  \node[anchor=west, font=\bfseries\large] at (1.2mm,8.8mm) {mod.c};
  \draw[gray!70] (1.2mm,6.5mm) -- (45.5mm,6.5mm);
  \node[anchor=north west, align=left, font=\ttfamily\footnotesize] at (1.2mm,5.5mm) {%
int mod\_run(void) \{ ... \}};
\end{scope}

\begin{scope}[shift={(modh.center)}]
  \coordinate (modhwest) at (0,6mm);

  \draw[codefiledraw]
    (0,0) -- (0,12mm) -- (44mm,12mm) -- (47mm,9mm) -- (47mm,0) -- cycle;
  \draw[codefiledraw] (44mm,12mm) -- (44mm,9mm) -- (47mm,9mm);

  \node[anchor=west, font=\bfseries\large] at (1.2mm,8.8mm) {mod.h};
  \draw[gray!70] (1.2mm,6.5mm) -- (45.5mm,6.5mm);
  \node[anchor=north west, align=left, font=\ttfamily\footnotesize] at (1.2mm,5.5mm) {%
int mod\_run(void);};
\end{scope}

\node[right=27mm of package, yshift=45mm] (criticsanchor) {};

\node[font=\bfseries, yshift=1mm] at (criticsanchor) (verificationtitle) {\stageverify};
\node[font=\bfseries] at ([yshift=-1mm]verificationtitle.south) (criticstitle) {\textit{Critics}};

\node[critic, below=1mm of criticstitle, xshift=-2mm] (compile) {\gcc};
\node[critic, below=1.8mm of compile] (framac) {\wpplugin};
\node[critic, below=1.4mm of framac] (cppcheck) {\cppchecker};
\node[critic, below=1.4mm of cppcheck] (vernfr) {\plugin};

\begin{scope}[on background layer]
\node[
    draw=black!40,
    dashed,
    rounded corners=3pt,
    line width=0.6pt,
    inner sep=1.2mm,
    fit=(verificationtitle) (criticstitle) (compile) (framac) (cppcheck) (vernfr)
] (criticsbox) {};
\end{scope}

\coordinate (agentsplit) at ([xshift=7mm]agent.east);

\draw[thick] (agent.east) -- (agentsplit);
\draw[->, thick] (agentsplit) |- (modcwest);
\draw[->, thick] (agentsplit) |- (modhwest);

\coordinate (ifacesplit) at ([xshift=3mm]iface.east);
\coordinate (formalsplit) at ([xshift=3mm]formal.east);
\coordinate (headersplit) at ([xshift=3mm]headers.east);

\coordinate (ifacejoin) at ([xshift=-1mm]modiswest);
\coordinate (formaljoin) at ([xshift=-1mm]fspecwest);
\coordinate (headerjoin) at ([xshift=-1mm]hdrswest);

\draw[thick] (iface.east) -- (ifacesplit);
\draw[thick] (formal.east) -- (formalsplit);
\draw[thick] (headers.east) -- (headersplit);

\draw[thick] (ifacesplit) |- (ifacejoin);
\draw[thick] (formalsplit) |- (formaljoin);
\draw[thick] (headersplit) |- (headerjoin);

\draw[->, thick] (ifacejoin) -- (modiswest);
\draw[->, thick] (formaljoin) -- (fspecwest);
\draw[->, thick] (headerjoin) -- (hdrswest);

\begin{scope}[on background layer]
\coordinate (implNW) at ([xshift=-3mm,yshift=5mm]package.north west);
\coordinate (implNE) at ([xshift=3mm,yshift=5mm]package.north east);
\coordinate (implE)  at ([xshift=3mm,yshift=-2mm]package.south east);
\coordinate (implSW) at ([xshift=-6mm,yshift=-22mm]agent.south west);
\coordinate (implW)  at ([xshift=-6mm,yshift=5mm]agent.north west);
\coordinate (implTopJoin) at ([xshift=3mm,yshift=-3mm]input.south east);

\draw[
    dashed,
    draw=black!40,
    line width=0.6pt,
    rounded corners=3pt
]
(implTopJoin) -- (implNW) -- (implNE) -- (implE) -- (implSW) -- (implW) -- cycle;
\end{scope}

\node[font=\bfseries, fill=white, inner sep=1pt]
    at ($(agent.south)!0.5!(package.south) + (21mm,84mm)$) {\stageimpl};

\draw[->, thick] (package.east) -- ++(4mm,0) |- (compile.west);

\coordinate (compilesplit) at ([yshift=-1.1mm]compile.south);
\coordinate (critictrunkNW) at ([xshift=-3mm]framac.west);
\coordinate (critictrunkSW) at ([xshift=-3mm]vernfr.west);

\draw[thick] (compile.south) -- (compilesplit);
\draw[thick] (compilesplit) -| (critictrunkNW);
\draw[thick] (critictrunkNW) -- (critictrunkSW);

\draw[->, thick] (critictrunkNW |- framac.west) -- (framac.west);
\draw[->, thick] (critictrunkNW |- cppcheck.west) -- (cppcheck.west);
\draw[->, thick] (critictrunkNW |- vernfr.west) -- (vernfr.west);

\node[
    outputparser,
    right=-28mm of vernfr,
    yshift=-18mm,
    minimum height=6mm,
    minimum width=38mm
] (parser) {Output parser};

\draw[thick] (compile.east) -- ++(4mm,0);
\draw[thick] (framac.east) -- ++(4mm,0);
\draw[thick] (cppcheck.east) -- ++(4mm,0);
\draw[thick] (vernfr.east) -- ++(4mm,0);

\draw[thick] ([xshift=4mm]compile.east) -- ([xshift=4mm]vernfr.east);
\draw[->, thick] ([xshift=4mm]cppcheck.east) -- ([xshift=4mm]cppcheck.east |- parser.north);

\node[
    logbox,
    below=2.5mm of parser,
    minimum width=30mm,
    minimum height=22mm,
] (report) {};

\node[font=\bfseries] at ([yshift=-3mm]report.north) {Report};

\draw[gray!70] ([xshift=2mm,yshift=-6mm]report.north west) --
               ([xshift=-2mm,yshift=-6mm]report.north east);

\node[anchor=north west] at ([xshift=0mm,yshift=-5mm]report.north west) {%
\begin{tabular}{@{}l@{}}
\scriptsize \gcc: \hspace{11mm}$\checkmark$ or $\times$\\
\scriptsize \wpplugin:\hspace{14.2mm}$\checkmark$ or $\times$\\
\scriptsize \cppchecker:\hspace{7.3mm} $\checkmark$ or $\times$\\
\scriptsize \plugin: \hspace{7.7mm}$\checkmark$ or $\times$
\end{tabular}
};

\coordinate (vleft)  at ([xshift=-10.5mm]report.south);
\coordinate (vright) at ([xshift=10.5mm]report.south);

\node[verdictnonverified, below=3mm of vleft] (notverified) {
\begin{minipage}[c][8mm][c]{8mm}
\centering
{\Huge $\times$}
\end{minipage}\\
\scriptsize Not verified
};

\node[verdictverified, below=3mm of vright] (verified) {
\begin{minipage}[c][8mm][c]{8mm}
\centering
\includegraphics[width=8mm]{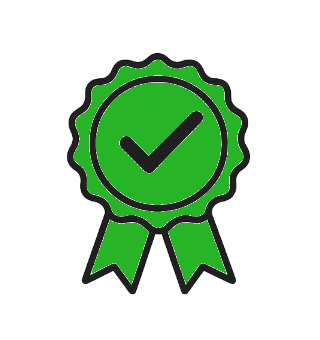}
\end{minipage}\\
\scriptsize Verified
};

\node[font=\bfseries, fill=white, inner sep=1pt]
    at ([yshift=4mm]parser.north) (outputtitle) {\stageoutput};

\begin{scope}[on background layer]
\node[
    draw=black!40,
    dashed,
    rounded corners=3pt,
    line width=0.6pt,
    inner sep=1.8mm,
    fit=(outputtitle) (parser) (report) (notverified) (verified)
] (outputbox) {};
\end{scope}

\draw[->, thick] (parser.south) -- (report.north);
\draw[->, thick] (vleft) -- (notverified.north);
\draw[->, thick] (vright) -- (verified.north);

\end{tikzpicture}%
}
\caption{Proposed \module development and verification workflow and \toolchain based on \plugin and \framac, where an untrusted agent generates embedded C code that is either verified by critics or fails to verify.
}
\label{fig:workflow}
\end{figure}
\vspace{8mm}

\autoref{fig:workflow} illustrates the proposed \toolchain for formal verification together with non-functional requirement verification based on \plugin (as described in \autoref{sec:implementation}). The workflow enables constructing a verified implementation of an embedded system C \module \textit{mod}. The colors in the figure indicate trust: blue elements represent trusted artifacts, while yellow elements represent untrusted artifacts and actors.

The workflow consists of four steps. In \stageinput, trusted artifacts specify the expected behavior and interface of the \module. In \stageimpl, an \agent generates a candidate \module in the form of a \cfile and a \hfile. In \stageverify, the \module implementation is validated by a set of trusted critics. In \stageoutput, the results are aggregated into a report and a verification outcome.

At step \stageinputnotitle, we specify the input needed to construct a verified implementation of a \module. An \agent is provided with the following information: 
\begin{itemize}
    \item \textbf{An \interfacespec}: A \module \iscontract, as defined in \autoref{sec:module-contracts}, defining entry-point functions, permitted external calls, and call-order constraints.
    \item \textbf{A natural language specification}: Describes the desired functional behavior of each entry-point function in the \iscontract in natural language.
    \item \textbf{A formal specification}: \acsl contracts for the entry-point functions from the \interfacespecs. These contracts must be consistent with the natural-language requirements to avoid contradictions. Additionally, \acsl contracts are provided for all external functions allowed by the \interfacespecs. The external contracts are necessary both for \wpplugin to verify the entry-point functions and for the \agent to correctly implement the module, as they specify the expected behavior of all permitted external calls.  
    \item \textbf{Header files}: All header files required for successful compilation. These provide macros, type aliases (typedefs), constants, and function declarations.
\end{itemize}

In \stageimplnotitle, the trusted inputs are provided to an untrusted \agent. The \agent may represent a human programmer or an automated system such as an LLM. It produces the implementation files \cfile and \hfile, which together form the candidate \module. 
The \cfile may also contain ACSL specifications that aid the verification, e.g., ACSL contracts for local helper functions or loop invariants.
Since the implementation is generated by the \agent, it is considered untrusted and must be verified. The generated files are combined with the trusted specifications to become the input to the next step.

In \stageverifynotitle, the \module package is analyzed by a set of trusted critics, each targeting complementary properties, including syntactic and functional correctness, coding-standard compliance, and interface-level constraints.
Compilation with \gcc is the first critic. It ensures that the program is syntactically correct, well-typed, and that all required declarations are available. We run \gcc as the first critic since the results of the remaining critics are not useful if the module does not compile.
After compilation, we run the remaining critics. The \framac plugin \wpplugin (see \autoref{sec:acsl_framac}) verifies functional correctness, i.e., that the entry-point functions implement their \acsl contracts. We run \wpplugin with the \wprteplugin option to ensure the absence of runtime errors.
Next, the \cppchecker critic performs static analysis with respect to the \misrac coding guidelines (see \autoref{sec:nfr}). 
Finally, the \plugin critic described in \autoref{sec:implementation} checks adherence to the non-functional rules defined in \autoref{sec:rules}. Since all critics are considered trusted, successful verification indicates that the implementation satisfies the checked properties.

In \stageoutputnotitle, the outputs from all critics are collected by an output parser, which produces a unified report summarizing the results. The final outcome consists of the \module package, the generated report, and a verification verdict. Based on the collected results, the \module is classified as verified or not verified. Since the critics are not complete, a negative result only indicates that correctness could not be established. However, for many programs, the critics are complete enough to reach a verdict and draw conclusions.

\section{Case Studies}
\label{sec:case_studies}
In this section, we present two case studies on ECU controller modules from \scania trucks, for which we have applied the \toolchain and workflow shown in \autoref{fig:workflow}. Before introducing the modules in detail, we describe the general approach we used for the case studies and the effort required.

\subsection{Specification and verification approach and effort}
To run the \toolchain on each module, we first wrote ACSL specifications and \iscontracts, based on existing requirement documents in natural language written by engineers. The ACSL specifications consist of function contracts for each entry-point function. 
We also wrote ACSL contracts for the external functions that are permitted to be called by the \iscontract. 
Both the ACSL contracts for the external functions and the entry-point functions could partially be reused from previous work~\cite{Ung2024,Sevenhuijsen2025a}. 

After writing the ACSL specifications and \iscontracts, we identified an existing \emph{reference implementation} of each module, corresponding to the latest available version of the code written by software engineers. We then ran steps (3)-(4) in \autoref{fig:workflow} for the reference implementations along with the ACSL specifications and \iscontracts that we wrote. We also conducted preliminary experiments for our \toolchain with LLM-based agents that implement the modules. The results for the reference implementations are presented in \autoref{sec:verification_results}, and the results for the LLM-based implementations are presented in \autoref{sec:llm_experiments}.

Since the modules are proprietary, we cannot provide their source code, ACSL specifications, or the full version of their respective natural language requirements. Instead, we give an overview of each module.

The total specification and verification effort amounts to approximately two person-weeks per case study for a verification engineer. This includes writing the ACSL specifications and the \iscontract, and ensuring that they are satisfied by the reference implementations. We estimate that writing the \iscontract required approximately one person-day per case study; the remaining time was dedicated to writing the functional ACSL specifications.

\subsection{Steering Fluid Level Detection (SFLD) case study}
The SFLD (Steering Fluid Level Detection) module diagnoses the steering fluid level in a power steering system for a vehicle. The role of a power steering system is to reduce the manual force needed by drivers to steer the vehicle, for which \emph{steering fluid} is a critical component. 
SFLD should monitor the steering fluid level and, if it drops below a certain threshold, a warning should be emitted to the driver. SFLD is safety-critical since the loss of power steering means that the vehicle becomes difficult to operate. The requirements document for SFLD consists of a summary of the intended functionality of the module, a control flow diagram, and seven requirements expressed in a mixture of natural language and pseudo-code. An important component described in the requirements document is a delay filter, similar to our simplified example in \autoref{fig:mod_good_example}. The description also details how the delay filter should be initialized upon vehicle startup.
\begin{wraptable}[10]{R}{0.40\textwidth}
\centering
 \caption{Size in LOC for SFLD.
 }
    {
    \begin{tabular}
    {lr}
    \toprule
         C implementation & 88
         \\
         Entry-point ACSL spec. & 127
         \\
         External ACSL spec. & 109
         \\
         \iscontract & 35
         \\
        \bottomrule
    \end{tabular}
    }
    \label{tab:sfld_metadata}
\end{wraptable}

\begin{figure}[tb]
    \centering
\begin{lstlisting}[language=ISContract, basicstyle=\scriptsize\ttfamily, backgroundcolor = \color{gray!10!white}]
module sfld {
  entry_functions: { void Sfld_create(void), void Sfld_10ms(void) }
  entry_order: { Sfld_create < Sfld_10ms }
  external_calls: {
    rteg.h: {
      void Rtdb_createBs(tBs sig), void Rtdb_createS32s(t32s sig),       
      tBs Rtdb_readBs(RTDBs sig), void Rtdb_writeBs(RTDBs sig, tBS bs), 
      void Rtdb_writeS32s(RTDBs sig, tS32S sig), ...  }
  }
  external_call_order: { Rtdb_createBs < Rtdb_WriteBs, ... }
}
\end{lstlisting}

    \caption{Excerpt of the \iscontract for SFLD.}
    \label{fig:sfld_iscontract}
\end{figure}

\autoref{tab:sfld_metadata} shows the size of the implementation and the contracts for SFLD in lines of code (LOC). Entry-point ACSL spec. refers to the ACSL specifications for the entry-point functions in SFLD and external ACSL spec. to the ACSL specifications for the external functions defined in the \iscontract. In \autoref{fig:sfld_iscontract}, we show an obfuscated excerpt of the \iscontract for SFLD. The types \lstinline!tBs!, \lstinline!t32s!, and \lstinline!RTDBs! are defined in an external module dedicated to defining custom Scania types.

The \iscontract for SFLD specifies that the module shall consist of two entry-point functions, \lstinline$Sfld_create$ and \lstinline$Sfld_10ms$. \lstinline$Sfld_create$ is assumed to be called at vehicle startup and \lstinline$Sfld_10ms$ is assumed to be called every 10~ms by a scheduler, hence, \lstinline$Sfld_create < Sfld_10ms$. 
The \iscontract also specifies 12 external functions together with four external call order constraints. The external function calls allow the module to interact with RTDB through the intermediate module RTEG, which in turn calls RTDB. The order constraints ensure that the RTDB variables are initialized before they are read from or written to.  

The reference implementation of SFLD consists of the two entry-point functions from the \iscontract, and no local functions. \lstinline$Sfld_create$ initializes the RTDB variables used by SFLD with calls to, e.g., \lstinline$Rtdb_createBs$. 
\lstinline$Sfld_10ms$ first reads the value of the oil-level sensor from RTDB and then implements the delay filter described in the requirements by incrementing or decrementing a timer that is stored in RTDB. If the timer reaches an upper threshold value, an oil-level warning is activated on the vehicle dashboard. If the timer reaches 0, the warning is deactivated. 
The implementation contains 17 call sites with calls to external modules, and uses one global variable with a static storage specifier, which serves as a Boolean flag to indicate if this is the first time the function is called since the last vehicle startup. 

\subsection{Safe Gearbox Maneuvering Module (SGMM) case study}
\begin{wraptable}[10]{R}{0.40\textwidth}
\centering
    \caption{Size in LOC for SGMM.
    }
    {
    \begin{tabular}
    {lr}
    \toprule
         C implementation & 138
         \\
         Entry-point ACSL spec. & 124
         \\
         External ACSL spec. & 58
         \\
         \iscontract & 28 
         \\
        \bottomrule
    \end{tabular}
    }
    \label{tab:sgmm_metadata}
\end{wraptable}
The SGMM (Safe Gearbox Maneuvering Module) is a controller for a gearbox in an electric vehicle. Specifically, it controls two magnetic valves, the high and the low valve, which in turn control the gearbox. When there is a risk for overspeed of the vehicle engine,  SGMM should block the magnetic valves to ensure that the gearbox is not activated.
SGMM is safety-critical since overspeeding of the engine may lead to fatal incidents. 
The requirements document for SGMM consists of a summary of the functionality of the module in natural language, a control flow diagram, and five requirements expressed as pseudo-code. The control flow diagram and the five requirements both express the conditions under which the module should ensure that the magnetic valves are blocked.

\begin{figure}[b]
    \centering
\begin{lstlisting}[language=ISContract, basicstyle=\scriptsize\ttfamily, backgroundcolor = \color{gray!10!white}]
module sgmm {
  entry_functions: { void Sgmm_10ms(void) }      
  external_calls: {
    aux.h: { void Aux_enterSafeSw(void), void Aux_exitSafeSw(void), ... },
    rtdb.h: {
        tB Rtdb_LowValve_read(void), tB Rtdb_HighValve_read(void),
        void Rtdb_LowValve_write(const tB val), 
        void Rtdb_HighValve_write(const tB val), ...
    },
    util.h: {void Util_registerEvent( void* moduleName, tU16 __LINE__ )}
  }
  external_call_order: { Aux_enterSafeSw <  Aux_exitSafeSw }
}

\end{lstlisting}
    \caption{Excerpt of the \iscontract for SGMM.}
    \label{fig:sgmm_iscontract}
\end{figure}

\autoref{tab:sgmm_metadata} shows the size of the implementation and the contracts in LOC. \autoref{fig:sgmm_iscontract} shows an obfuscated excerpt of the \iscontract. As for SFLD, the type \lstinline!tB! is defined in the external module for type definitions. The \iscontract for SGMM specifies 13 external functions and one external call order constraint. There is no initialization function in SGMM, since it uses a version of RTDB where initialization is not delegated to the application modules. Aside from reading and writing to RTDB, the implementation calls functions that control memory-safety aspects in the ECU. For example, the call to \lstinline$Auxc_enterSafeSw$ activates runtime memory-safety checks in the ECU.

The implementation of SGMM consists of the single entry-point function \lstinline$Sgmm_10ms$, expected to be called repeatedly by the scheduler every 10~ms. The implementation first reads input values from RTDB, and then checks if any of the conditions for blocking the valves are satisfied. If so, it blocks the valves by writing to the corresponding RTDB output variables. \lstinline$Sgmm_10ms$ is implemented using several local functions and 17 calls to the external modules. There are no global variables in the implementation.

\subsection{Verification results}
\label{sec:verification_results}
This section reports the results for the case studies using the \toolchain described in \autoref{sec:workflow}, which we implemented in Python. 
First, we run the compilation using \gcc version 16.1 with default compilation parameters. Then, we perform deductive verification using \framac version 32.0 and the \wpplugin plugin with the \wprteplugin option. Further static analysis is carried out using \cppchecker version 2.21 with the \misrac:2012 addon. Lastly, we use \plugin for the non-functional verification. Each critic reports its result to a shared aggregation step, which produces a unified report and determines the final verification outcome.

\begin{table}[tb]
\centering
\caption{Verification results for the SGMM and SFLD reference implementations. For \framac{} WP, $(p/t)$ denotes $p$ proven proof obligations out of $t$ generated obligations. For \cppchecker, \misrac violations are described inside parentheses.}
\label{tab:verification-results}
\footnotesize
\setlength{\tabcolsep}{8pt}
\begin{tabular*}{\textwidth}{@{\extracolsep{\fill}} l l r l r @{}}

& \multicolumn{2}{c}{SGMM} & \multicolumn{2}{c}{SFLD} \\
\cmidrule(lr){2-3} \cmidrule(lr){4-5}
Critic & Result & Time [s] & Result & Time [s] \\
\midrule

\gcc
& $\checkmark$ & 0.4
& $\checkmark$ & 0.3 \\

\wpplugin
& $\checkmark$ (86/86)
& 14.8
& $\checkmark$ (80/80)
& 8.4 \\

\cppchecker
& $\times$ (6 required)
& 58.9
& $\times$ (4 required, 2 advisory)
& 20.1 \\

\plugin 
& $\checkmark$
& 13.2
& $\checkmark$
& 12.0 \\

\midrule
Total
& & 87.3
& & 40.7 \\

\bottomrule
\end{tabular*}
\end{table}

\autoref{tab:verification-results} summarizes the verification results for the reference implementations of the two case studies. The Critic column lists the analyses applied to each \module. The remaining columns are grouped by case study. For each case study, the Result column reports the verdict, and the Time column reports the corresponding analysis time in seconds. Reported times are CPU time measured using the Unix \texttt{time} command as the sum of user and system time. All experiments were performed in a virtualized Linux environment (WSL2) running on a machine with an Intel Core i7-1365U CPU (6 cores, 12 threads) and 16 GB RAM.

As shown in the table, both SGMM and SFLD compile successfully using \gcc, all proof obligations generated by \wpplugin are proven, and no violations are reported by \plugin. 

For SFLD, no \misrac violations are reported when restricting the analysis to the module implementation itself, i.e., when we exclude included header files. For SGMM, however, \cppchecker reports six required \misrac violations in the module implementation itself. More specifically, two violations concern assignments between expressions of incompatible or narrower essential type categories (Rule 10.3), and four violations concern missing compound statements as bodies of selection statements (Rule 15.6).
When considering the full compilation unit, additional \cppchecker violations are also reported in included external library headers. For SFLD, these consist of four required and two advisory violations, primarily related to type consistency in expressions and scope-related recommendations. We discuss \cppchecker violations more generally in \autoref{sec:discussion}. %

The analysis time varies by an order of magnitude between the critics. The critic \cppchecker is the most expensive one in both case studies, with CPU time in the order of tens of seconds. \wpplugin, \plugin, and \gcc are significantly faster, with the former two requiring around ten seconds and \gcc less than a second. These analysis time orderings are consistent for both SGMM and SFLD. 

\subsection{Preliminary LLM experiments}
\label{sec:llm_experiments}
So far, our case studies have considered implementations by software engineer (human) \agents. However, the proposed \toolchain is agnostic to the origin of the code and also applies to implementations generated by LLM-based \agents. To explore this, we conducted preliminary experiments using two LLMs: Claude Sonnet 4.6 \cite{anthropic_claude_sonnet_46_system_card_2026}, accessed through AWS Bedrock \cite{aws_bedrock_2026}, and GPT-4o \cite{openai2024gpt4o} accessed through the OpenAI API.

In these experiments, we prompted the models with the natural-language specification, the interface specification, the required header files, and the relevant type definitions. We also imposed explicit constraints on the expected output format. In particular, the models were instructed to generate exactly one \cfile and one \hfile and to implement the entry-point functions from the interface specification. The prompt did not include the non-functional rules defined in \autoref{sec:rules}; instead, it provided information on the syntax and semantics of the \iscontracts.
For each generated \hfile and \cfile pair, we compiled them with \gcc since, as discussed in \autoref{sec:workflow}, successful compilation with \gcc is a baseline for the remaining critics. If compilation succeeded, we continued the \wpplugin critic, followed by the \cppchecker critic for \misrac. Lastly, the \plugin critic checks non-functional requirements for the rules defined in \autoref{sec:rules}.

Table~\ref{tab:llm_prelim_results} presents a representative subset of the conducted experiments, including both runs with the interface specification and runs in which the \iscontract was omitted from the input.
The LLM column identifies the model used to generate the implementation. The Module column identifies the generated module. The Interface column indicates whether the interface specification was included in the prompt. The LOC column reports the size in lines of code of the generated C implementation, and the Time column reports the generation time in seconds. The remaining columns report the outcomes of the critics. The functional verification with \wpplugin uses a global timeout of 600 seconds and a timeout of 2 seconds per proof obligation. We chose this timeout since initial experiments indicated that it was sufficient to verify correct implementations for the two case studies. 

\begin{table*}[t]
\centering
\caption{Representative preliminary LLM experiments on SGMM and SFLD. Each row reports one generated implementation. The column Time shows the time in seconds for LLM generation of the implementation. In the critics, $\checkmark$~$t$ denotes that the critic accepted the implementation in $t$ seconds, while $\times$~$t$ denotes rejection or unsuccessful completion after $t$ seconds.}
\label{tab:llm_prelim_results}
\footnotesize
\setlength{\tabcolsep}{4pt}
\begin{tabular*}{\textwidth}{@{\extracolsep{\fill}} l l l r r c c c c @{}}
\toprule
LLM & Module & Interface & LOC & Time & GCC & WP & \cppchecker & \plugin \\
\midrule
Sonnet 4.6 & SGMM & Yes &  93 & 105.1 & $\checkmark$ 1.7 & $\checkmark$ 8.6   & $\checkmark$  57.0  & $\checkmark$  14.3  \\
Sonnet 4.6 & SGMM & No  &  94 &  96.7 & $\checkmark$  1.6  & $\checkmark$  7.8    & $\checkmark$  50.8  & $\checkmark$  14.4  \\
GPT-4o     & SGMM & Yes &  37 & 120.8 & $\checkmark$  0.5  & $\checkmark$  17.6   & $\times$  62.4      & $\checkmark$  15.8  \\
Sonnet 4.6 & SFLD & Yes & 179 & 661.1 & $\checkmark$  1.2  & $\times$  600.0      & $\times$  13.1      & $\checkmark$  15.0  \\
Sonnet 4.6 & SFLD & No  & 183 & 655.9 & $\checkmark$  0.4  & $\times$  600.0      & $\times$  12.2      & $\checkmark$  12.5  \\
GPT-4o     & SFLD & Yes &  83 & 668.0 & $\checkmark$  1.7  & $\times$  600.0      & $\times$  21.3      & $\checkmark$  21.4  \\
\bottomrule
\end{tabular*}
\end{table*}

The results show that both models produced outputs in the requested format and, in all reported runs, compiled successfully. This matches the reference implementations, which also compile successfully. For SGMM, Sonnet 4.6 produced implementations that passed all critic columns both with and without the interface specification in the input. This differs from the SGMM reference implementation, which passed compilation, \framac{} WP, and \plugin, but was classified as not verified because \cppchecker reported required \misrac violations. GPT-4o also produced a compilable SGMM implementation and passed the \wpplugin and \plugin critics, but failed the \cppchecker critic. Its verification outcome is therefore closer to the SGMM reference implementation, although the underlying reported violations are not the same.

For SFLD, both Sonnet 4.6 and GPT-4o produced implementations that compiled and passed the \plugin critic, but functional verification remained more difficult, as \wpplugin timed out. This contrasts with the SFLD reference implementation, for which all \wpplugin proof obligations were discharged. We therefore classified the generated SFLD implementations as not verified. In addition, all reported SFLD runs failed the \cppchecker critic. This is consistent with the reference implementation when considering the full compilation unit, where \cppchecker also reports \misrac violations, although no \misrac violations are reported for the SFLD module implementation itself when included headers are excluded.

We also conducted experiments in which we omitted the \iscontract from the input. In several cases, the models still generated code that adhered to the intended interface. In these cases, only the \iscontract was omitted, while the external header files still provided substantial context to the LLM. We suspect that these header files provide enough contextual information for the model to infer which RTDB functions to use, since the function names often describe their intended roles. However, the \iscontract remains necessary to formally verify that the interface behavior of the generated code is correct.

Overall, these experiments suggest that generated implementations can reach the same basic compilation baseline as the reference implementations, and, in some cases, also satisfy the checked non-functional properties. However, full functional verification remains more sensitive to the complexity of the module and the completeness of the specifications. This is most obvious for SFLD, where the reference implementation is functionally verified but the generated implementations are not.

\section{Discussion}
\label{sec:discussion}
In this paper, we consider functional verification of safety-critical code a baseline on which to improve by specifying and verifying additional properties. Yet, simply reaching this baseline using \acsl and \framac for our case study modules took considerable effort, both in absolute terms and in comparison to writing and verifying IS contracts. For example, we used ghost variables to model RTDB, which required significant work to set up. The two modules in the case study also use different versions of RTDB, requiring us to adapt the model for the respective case study. Moreover,
for the SFLD case study in particular, we struggled to write the ACSL specifications due to the poor quality of the natural language requirements. The natural language specifications also come with a set of implicit assumptions on how code is written for Scania trucks. For example, Scania RTDB variables maintain a \emph{status} field for which the interpretation follows some Scania coding conventions. While we were able to study the reference implementation to resolve ambiguities and incompleteness, LLM agents may not have access to such implementations when prompted to generate modules. 
High quality requirements, both in natural and formal languages, are thus likely to be important in workflows similar to ours when relying on LLM agents.

When formulating the interface specifications, a crucial aspect was to identify the necessary external function calls to implement the two modules. In this case, we could harvest call sites from the reference implementations. When no reference implementation exists, this will be a significantly more difficult task, which will require the company to maintain proper documentation of the module interfaces. For the two modules we studied, however, the external calls are mainly to communicate with RTDB. For such cases, simply specifying the input and output RTDB variables per module will therefore naturally induce a large part of the interface specification. 

In the case studies, the module reference implementations were classified as not verified due to violations reported by \cppchecker. However, these violations do not necessarily indicate incorrect or unsafe behavior. Instead, we believe they reflect a mismatch between the standard \misrac rules enforced by \cppchecker and the project and company-specific adaptations of these rules used in \scania production code. In practice, such adaptations allow certain constructs that are disallowed or discouraged by the standard \misrac guidelines. As a result, the violations should be interpreted as deviations from strict \misrac compliance rather than violations of functional correctness or module-level requirements. 

In our \toolchain, we rely on \wpplugin for functional verification. Typically, the verification with \wpplugin is \emph{modular} and therefore requires that all local helper functions are annotated with ACSL function contracts, and similarly all loops with loop invariants. For smaller modules, it may be possible to instead use function inlining and loop unrolling, but Knüpper et al.~\cite{knupper18inlining} have shown that modular approaches are required for deductive verification to scale. In the \toolchain, we assume that the agent accompanies the implementation with contracts for the helper functions. In earlier work, we have developed the 
\autodeduct toolchain~\cite{Amilon2025,Amilon2024,alshnakat2020} for automated verification of industrial embedded systems software. A key part of \autodeduct is automatic \emph{contract inference} for helper functions. We envision \plugin to be integrated into \autodeduct in future releases, which would eliminate the need for the agent to annotate local helper functions and loops.

\section{Related Work}
Numerous approaches and tools permit specification and verification of non-functional properties. For C, the ACSL language offers specification of memory safety and termination, which can be verified by \wpplugin~\cite{wp_manual}. The \metacsl plugin~\cite{metacsl} of \framac extends \acsl with the Hilare language for \emph{meta properties}. \metacsl transforms the Hilare meta properties into (equivalent) regular ACSL specifications that can be verified using \wpplugin or other tools. Hilare allows for specifying properties related to function calls. In particular, we could have used Hilare and \metacsl to specify and verify the permitted function calls in a given module (rule~\ref{cfr:ext_calls}). However, we opted not to use Hilare, since it does not support specifying call order sequences  (rule~\ref{cfr:ext_calls_order}). 

Several existing formal specification languages consider control flow, such as CVPP~\cite{huisman11} for Java and the \framac \aorai plugin~\cite{aorai} for C. Both CVPP and \aorai use an automata language to specify allowed sequences of function calls and then verify that all executions of a program adhere to the automata. Another approach by Soleimanifard~et~al.~\cite{Soleimanifard2015} uses temporal logic annotations in programs to verify procedure-level control flow, which allows verification of, e.g., certain interleavings of function calls.
Compared to these approaches, the control flow side of our contract language is simpler and aimed at practical use in verification of embedded C code, but could straightforwardly be extended to support expressiveness equivalent to a class of finite automata.

Regarding contract languages for non-functional requirements more generally, Loques et al.~\cite{loques04} develop a contract-based framework for developing modular software that adheres to non-functional requirements, and Lopez et al.~\cite{lopez25} extend a contract language with properties such as execution time and memory consumption.
In contrast, our work is specific to module interface requirements of embedded C code related to control flow and data flow.

Recent work integrates LLMs with formal verification or symbolic reasoning tools to assess correctness properties of generated programs~\cite{patil24aisola,mukherjee_towards_2024,Slama_2024}. For C code, Sevenhuijsen et al.~\cite{Sevenhuijsen2025a} present a workflow in which an LLM generates candidate implementations that are iteratively refined using feedback from the compiler and verification tool. Patil et al.~\cite{patil24aisola} and Sevenhuijsen et al.~\cite{Sevenhuijsen2025b} evaluate the generation of automotive embedded C code from natural language and ACSL specifications, showing that compilable and partially or fully verifiable implementations can be obtained. While these approaches demonstrate the feasibility of combining LLM generation with verification, they target functional properties and do not explicitly address non-functional requirements such as control-flow restrictions, interface conformance, and inter-module interaction.
 Regarding non-functional properties of LLM-generated code, Sun et al.~\cite{Sun2025} study high-level non-functional quality characteristics, while Lin et al.~\cite{lin2025robunfr} show that enforcing non-functional properties may impact functional correctness. 

\section{Conclusion}

We presented principles and rules for capturing non-functional requirements on control flow and data flow in embedded systems C code, operationalized in a checker built on the \framac framework~\cite{amilon2026vernfr}. Use of our checker is complementary to functional verification of \acsl properties and \misrac conformance checking. The checker relies on a module-level contract language that extends a bridge from ACSL function contracts towards system level contracts, e.g., expressing end-to-end safety requirements of trucks.

Heavyweight formal methods such as interactive theorem proving can, in principle, support trustworthy modular reasoning about system level properties, including non-functional properties, for embedded systems with software in C~\cite{Beringer2021,Carbonneaux2014}. However, they remain far from engineering practice and industry productivity expectations. By considering more lightweight methods and languages for non-functional verification, we believe safety benefits and LLM productivity increases are more likely to accrue in industrial applications.

Our case studies suggest that functional specification and verification remains a bottleneck in achieving trustworthy safety-critical software written in C, but can be used to pivot towards practical system level verification of a wider range of properties via the module abstraction. 

\begin{credits}
\subsubsection{\ackname} The authors thank Gustav Ung, Minal Suresh Patil, and Dilian Gurov for discussions on this work, and the anonymous reviewers for their comments. This work was supported by Vinnova through the FormAI project and by Digital Futures at KTH through the Open Sandbox project.
\end{credits}

%
%
%
\bibliographystyle{splncs04}
\bibliography{ISoLA26/bib}

\end{document}